\documentclass[a4paper,aps,pra,twocolumn,
preprintnumbers]{revtex4}
\usepackage{graphics,graphicx,epsfig}

\usepackage[centertags]{amsmath}
\usepackage{amsfonts}
\usepackage{amssymb}

\usepackage[cp1251]{inputenc}
\usepackage[russianb]{babel}

\usepackage[centertags]{amsmath}
\usepackage{amsfonts}
\usepackage{amssymb}
\usepackage{amsthm}
\newcommand{\beq}{\begin{equation}}
\newcommand{\eeq}{\end{equation}}
\newcommand{\beqa}{\begin{eqnarray}}
\newcommand{\eeqa}{\end{eqnarray}}

\newcommand{\ket} [1] {\vert #1 \rangle}
\newcommand{\bra} [1] {\langle #1 \vert}

\newcommand{\mean}[1]{\langle #1 \rangle}

\begin{document}
\sloppy

\begin{abstract}

A method based on integrals of motion for collective processes has
been introduced to achieve physical schemes in which one of the
systems is insensitive to interaction. Decoherence-free quantum
channels that allow sending any state of light, particulary the Fock
states, through an absorbing medium are considered as an example.
\end{abstract}

\title{Quantum channel for light based on integrals of motion}


\author{V.N. Gorbachev, A.I. Trubilko}

\affiliation {Laboratory for Quantum Information $\&$ Computation,
AeroSpace University, St.-Petersburg 190000, Bolshaia Morskaia 67,
Russia}

\maketitle

\section{Introduction}
Integrals of motion  may result in conserving several properties of
the interacting systems, that are interesting for applications.
\\
In the quantum informational processing the problem of integrity of
the state of the physical system is significant thanks to
decoherence, which destroys the state because of the irreversible
interaction with environment. One of solutions of the problem is to
refer to decoherence-free subspases consisting of states which have
immunity to interaction and store their integrity. First
decoherence-free subspaces have been introduced by Zanardi \cite{1}
by considering  interaction between $N$ two-level atoms and the
multimode electromagnetic field. For this problem the subspace of
atomic wave functions that are annihilated by interaction
Hamiltonian has been found. For N = 2, the subspace includes only
one wave function $\Psi^{-}$, which is one of the Bell states,
antisymmetric with respect to the permutations of particles. As it
was shown by Basharov  \cite{2} this state has immunity to decay in
the collective thermostat. Some interesting examples of
decoherence-free states for the spin-spin and spin-boson
interactions have been considered by many authors (see, for example
\cite{3,4}). Weinfurter et al have demonstrated experimentally a
scheme of decoherence-free communication based on the four-photon
polarized states of light \cite{5}.
\\
To achieve decoherence-free spaces integrals of motion can be used.
In the various processes of interactions between light and resonant
and transparent media integrals of motions may result in conserving
quantum correlations between modes \cite{6,7}, that is a basis, for
example, of amplification of EPR pair of continuous variables
\cite{8}.
\\
The main aim of our work is to study decoherence-free communication
using integrals of motions. We show, that existence of integral of
motions can establish such type of interaction between two systems
that one of these systems does not change. Then we find a
decoherence-free space including all states of the system in
contrast to decoherence-free-subspace. For example we consider the
set of schemes involving two modes of light and absorbing atoms.
When there is only one of the modes  it is absorbed by atoms when
propagating through the medium. However despite of absorbtion it can
be reproduced at the output if the second mode and some additional
optical elements are introduced in the schemes. These schemes can be
considered as a decoherence free cannel for sending any quantum
state of light.  We focus on a particular problem of sending light
in the Fock state through the absorbing medium. For this case the
decoherence-free channel can be achieved with the help of two
non-absorbing beamsplitters and a mode in the coherence state. We
consider the Fock states of light, particulary a single-photon state
as they are interesting for many problems. Indeed in quantum
computation many operations can be implemented by linear optics
elements if the single-photon states are used.  In the KLM model
\cite{9} based on linear optics logical variables $0$ and $1$ are
encoded by the two-mode Fock states $\ket{01}, \ket{10}$. These are
non-classical states of light and they have quantum correlations
which have to be protected against to decoherence for successful
computations.
\\
The paper is organized as follows. First we consider a general
conditions of existence of integral of motions and methods to
achieve the schemes including two modes, one of which is insensitive
to interaction. Then we discuss the optical schemes for sending
light through absorbing medium presented by two-level atoms. As an
example a problem of sending the Fock state of light is considered.
The task can be accomplished when the Fock state is mixed with a
mode in coherent state. Such state obtained after mixing has been
demonstrated experimentally by Lvovsky \cite{10}, its properties is
discussed in Appendix.

\section{Integral of motion}

Let $H_{AB}$ be the Hamiltonian of two interacting system $A$ and
$B$ and $Z=Z(A,B)$ be an observable, which can depend on the
variables of both systems. The evolution of operator $Z$ is given by
\begin{eqnarray}
  Z'=T^{\dagger}_{AB}ZT_{AB},
\end{eqnarray}
where $T_{AB}=\exp(-i\hbar^{-1}H_{AB}t)$ and we assume for
simplicity that the Hamiltonian $H_{AB}$ is time independent. If
\begin{eqnarray}
\label{0001}
 [T_{AB}; Z]=0,
\end{eqnarray}
then $Z$ is an integral of motion. Condition (\ref{0001}) can be
achieved in various cases. Let the evolution operator be independent
from the variables of one of the systems, say, $B$:
\begin{eqnarray}
\label{0002} T_{AB}=T_{A}\otimes 1_{B}.
\end{eqnarray}
Then it is clear that any observable $Z(B)$ is integral of motion.
\\
A nontrivial solution of Eq. (\ref{0002}) can be found by means of
unitary transformations. Let $U(A,B)$ be a unitary operator
depending on the variables of $A$ and $B$. Then
\begin{eqnarray}
  U^{\dagger}(A,B)(S(A)\otimes 1_{B})U(A,B)=S(A,B)
\end{eqnarray}
for any operator $S(A)$ and one finds Eq. (\ref{0002})
\begin{eqnarray}
\label{00033} T_{AB}=U(A,B)S(A,B)U^{\dagger}(A,B) =S(A)\otimes
1_{B}.
\end{eqnarray}
In accordance with (\ref{0002}) the evolution of the density matrix
of two systems has the form
\begin{eqnarray}
\label{0004}
 \rho'_{AB}=T_{AB}(\rho_{A}\otimes\rho_{B})T^{\dagger}_{AB}=\rho'_{A}
 \otimes\rho_{B}.
\end{eqnarray}
Eq. (\ref{0004}) tells that the system $B$ described by the density
matrix $\rho_{B}$ is unchanged during interaction if there is
integral of motion. Note that this result is valid for any state of
$B$. In contrast to our case  Zanardi et al used another method and
found a finite set of states $\{\Psi_{B}\}$ which are annihilated by
the Hamiltinian of interaction: $H_{AB}\Psi_{B}=0$.
\\
Let the unitary operators $R_{A}$ и $N_{B}$ depend on the variables
of $A$ and $B$ respectively. Then a simple transformation of Eq.
(\ref{0002}) aries
\begin{eqnarray}
\label{09}
  (R_{A}\otimes N_{B})^{\dagger}T_{AB}
  (R_{A}\otimes N_{B})=T'_{A}\otimes 1_{B},
\end{eqnarray}
which holds the above properties. The obtained Eq. (\ref{09})
provides consideration of various physical schemes.
\\
Eq. (\ref{00033})  has a simple interpretation if the operator
$S(A,B)$ is unitary and hence can be associated with an Hamiltonian
of interaction $V_{AB}$: $S(A,B)=\exp(-iV_{AB}t)$. In this case the
next property aries. The existence of integral of motion allows to
choose a form of interaction and to construct a scheme including
three sequential transformations $U(A,B)S(A,B)U^{\dagger}(A,B)$,
which keep the system $B$ unchanged. Then using this scheme and
unitary transformation (\ref{09}) as initial resource one can
construct a set of unitary equivalent schemes of the forme
$U'(A,B)S'(A,B)U'^{\dagger}(A,B)$, where $X'=(R_{A}\otimes
N_{B})^{\dagger}X(R_{A}\otimes N_{B})$ and $X=U(A,B), S(A,B)$, which
accomplish the same task.
\\
As an example consider the Bogolubov transformation of two boson
operators $a$ and $b$: $U^{\dagger}aU=ca+sb, U^{\dagger}bU=-sa+cb$,
where $c^{2}+s^{2}=1$. Then $ S(ca+sb)=U^{\dagger}S(a)U,$ where $S$
is a unitary operator, that can describe the physical interaction
between two modes that is represented by the collective operator
$ca+sb$. It is clear that $T_{ab}= US(ca+sb)U^{\dagger}=S(a)\otimes
1_{b}$. Then we use (\ref{09}). Let $R_{A}$ be the phase shift
operator: $R_{a}: a\to a\exp(i\mu)$, where $\mu$ is a real number
and $N_{B}=1$, then $T_{ab}\to
T'_{ab}=U'S'(ca\exp(i\mu)+sb)U'^{\dagger}=S(a)\otimes 1_{b}$, where
$U'=R_{a}^{\dagger}UR_{a}$.
\\
The general properties presented above can be realized for
particular physical processes.

\section{Optical schemes}

We consider two optical schemes with interaction between two el.
modes $a$ and $b$ and two-level absorbing atoms. In both schemes
frequency of the mode $b$ is resonance to the frequency of atomic
transition hence this mode may be absorbed. However the mode $b$
becomes  insensitive to absorbtion if there is a second mode $a$. We
consider two types of interactions: 1/ single photon absorbtion, 2/
two-photon and Raman type interactions.
\\
1/ Let Hamiltonian of atoms and field has the form
\begin{eqnarray}
\label{001} \label{21} H=H_{P}+H_{F}+V,
\end{eqnarray}
where $H_{P}$ and
$H_{F}=\hbar\omega_{a}a^{\dagger}a+\hbar\omega_{b}b^{\dagger}b$ are
Hamiltonians of free atoms and field, which describs by its
operators of creation and annihilation of photons of the modes
 $a^{\dagger}, a$ and
$b^{\dagger}, b$, with frequencies $\omega_{a},\omega_{b}$. In
dipole and rotating wave approximations operator of interaction
between atoms and field takes the form
\begin{eqnarray}
\label{22} V=i\hbar[S_{10}(ga+fb)-S_{01}(ga+fb)^{\dagger}],
\end{eqnarray}
where  $S_{xy}=\sum_{m}s_{xy}(m)$, $s_{xy}(m)=\ket{x}_{m}\bra{y}$ is
a single-atom operator, $x,y=0,1$, $\ket{0}_{m}$, $\ket{1}_{m}$ are
lower and upper level of atom,  $g,f$ are coupling constants which
are assume to be real.
\\
Introduce new variables
\begin{eqnarray}
\label{002} 
 r=(ga+fb)G^{-1}, \tau=(fa-gb)G^{-1},
\end{eqnarray}
where $G=\sqrt{g^{2}+f^{2}}$. Then the Hamiltonian of interaction
(\ref{22}) depends only on $r$: $ V=i\hbar
G[S_{10}r-S_{01}r^{\dagger}],$ and equation of motion for $\tau$ has
the form
\begin{eqnarray}
\label{221} \frac{\partial}{\partial t}
\tau=i\hbar^{-1}[H,\tau]=-i(f\omega_{a}a-g\omega_{b}b)/G.
\end{eqnarray}
It follows, that for the equal frequencies
$\omega_{a}=\omega_{b}=\omega$  the linear combination of operators
$\tau=(fa-gb)G^{-1}$ is integral of motion because of its evolution
redices to the multiplication by a phase factor $\exp(-i\omega t)$.
\\
Next points may be made about it. First, the found integral remains
unchanged when atomic relaxation is taken into account by including,
for example, in the Hamiltonian (\ref{001}) an interaction between
atoms and thermostat. Second, any function
$\varphi(\tau,\tau^{\dagger})$ is also integral of motion.
\\
The presented model can describe the propagation of waves with
identical frequencies resonant to the atomic transition. Fot this
case, two waves can be distinguished by its polarization or
direction of its propagation and simultaneously interact with the
same atomic transition. These requirements are satisfied for, e.g.,
a broadband absorber that is insensitive to the light polarization.
Further, we will consider absorption for definiteness. The unitary
transformation $U$  given by Eq. (\ref{002}) selects the linear
combination of the modes that does not interact with atoms. The
transformation $U$ can be realized by means of either a nonabsorbing
beamsplitter   or a parametric down convertor with the classical
pumping wave. Both physical systems are described by the effective
Hamiltonian
\begin{eqnarray}
 \label{12345}
 \upsilon=i\hbar k(a^{\dagger}b-ab^{\dagger}),
\end{eqnarray}
where the modes $a$ and $b$ have the same frequencies and
polarization for the case of the beamsplitter  and they can have
different frequencies and polarization in the case of the parametric
down convertor.
\\
For definiteness we will consider the beamsplitter, then one finds
$U=\exp(-i\hbar^{-1}\upsilon t)$. Now introduce atomic relaxation by
adding to the Hamiltonian $H$  the operator of the interaction
between atoms and thermostat $V_{E}$ and  Hamiltonian of the free
thermostat $H_{E}$: $H\to H + H_{E} + V_{E}$. For simplicity, we
consider only the resonance interaction between the modes $a$, $b$
and atoms assuming that
\begin{eqnarray}
  \omega_{a}=\omega_{b}=\omega_{0},
\end{eqnarray}
where $\omega_{0}$ is the atomic transition frequency. Under this
conditions we find
\begin{eqnarray}
\label{000222} \nonumber
S(ga+fb)=&&\\
\nonumber
T\exp\{-i\hbar^{-1}\int_{0}^{t} dt'[V(ga+fb)+V_{E}(t')]\}&&\\
\nonumber
 =
U^{\dagger}T\exp\{-i\hbar^{-1}\int_{0}^{t}dt'[V(a)+ V_{E}(t')]\}U
&&\\
=U^{\dagger}S(a)U,&&
\end{eqnarray}
where $T$ is the time-order operator and we used the relation
$U^{\dagger}aU=(ga+fb)G^{-1}$ together with $[U; V_{E}]=0$. Equation
(\ref{00033}) immediately follows from this relation.
\\
Equation (\ref{000222}) enables to construct an optical scheme with
two modes and atomic medium, where one of the modes is insensitive
to absorption. To this end, two beamsplitters  are placed in front
of the absorber and behind it. The scheme operates as follows.
Let
the beamsplitters  have the same transmittance and reflectance given
by coefficients $c$ and $s$, where $c=f/G,s=g/G$. Let $a'$ and $,b'$
be the modes at the input of the absorbing medium and
 $a'', b''$  be these modes at the output of the absorber. Let
 the linear combination  $\tau=(fa'-gb')/G$ be integral of motion,
 then $\tau=(fa''-gb'')/G$. If modes $a''$ и $b''$  are mixed by the
beamsplitter placed behind the absorber, then one finds the integral
of motion $(ca''-sb'')/G=b_{out}$ at one of the outputs of the
beamsplitter, that is output of the scheme. It is convenient  to
consider the transformation of input modes $a',b'$ in the inverse
order. Let they be mixed at the beamsplitter placed in front of the
absorber. In this case one finds integral of motion
$(ca'-sb')/G=b_{in}$ at one of the outputs of the bemsplitter. But
this output is one of the inputs of the scheme, then
$b_{in}=\tau=b_{out}$. This means, that the mode $b$ is reproduced
at the end of the scheme being insensitive to absorbtion.
\\
2/ Now consider a set of the unitary equivqlent schemes obtained by
the the transformation (\ref{09}), where interactions with classical
waves are introduced. Let in Eq. (\ref{09}) operator $N_{B}=1$, and
$R_{A}$ is  phase shift operator $R_{a}$ of the mode $a$: $a\to
a\exp(i\epsilon\Omega t)$, where $\epsilon=\pm 1$ and $\Omega
>0$. If $\epsilon=0$, then one finds the previous case. All
resources of the new schemes are obtained by the unitary
transformation $R_{a}$ of the initial resources, but the
transformation leads to another physical processes. In particular,
Hamiltonian of interaction (\ref{22}) takes the form
\begin{eqnarray}
\label{22222} V'=i\hbar[S_{10}(gae^{\epsilon i\Omega
t}+fb)-S_{01}(ga^{\dagger}e^{-\epsilon i\Omega t}+fb^{\dagger})].
\end{eqnarray}
If $\epsilon=\pm 1$ the Hamiltonian describes the nondegenerate two
photon absorbtion and Raman type interaction between the mode $a$, a
strong  classical wave at frequency $\Omega$ and atoms :
\begin{eqnarray}
 \nonumber
  \omega_{a}+\Omega=\omega_{b}=\omega_{0},&&\\
  \omega_{a}-\Omega=\omega_{b}=\omega_{0}.&&
\end{eqnarray}
From these relations  new integral of motion
$\tau'=(fa\exp(i\epsilon\Omega) t-gb)/G$ follows under the
conditions that $\omega_{a}-\epsilon\Omega=\omega_{b}$, when
Hamiltonian $\upsilon$ given by (\ref{12345}) is replaced to
\begin{eqnarray}
 \label{123456}
 \upsilon'=i\hbar k(a^{\dagger}b\exp(-i\epsilon\Omega t)-
 ab^{\dagger}\exp(i\epsilon\Omega t)).
\end{eqnarray}
This effective Hamiltonian describes the three-photon parametric
process of up frequency conversion
$\omega_{a}-\epsilon\Omega=\omega_{b}$  with the classical pumping
wave at frequency $\Omega$. As a result, to transmit the mode $b$
through the absorbing medium in the obtained scheme it needs to mix
it with the mode $a$ in the parametric convertor, then to guide both
modes to the absorber, and to separate the mode $b$ using the second
parametric convertor.

\section{The quantum channel for the Fock states of light}

As an example consider a problem of sending a Fock state through an
absorber. According to Eqs. (\ref{00033}) and (\ref{22}) the task
can be accomplished by the scheme that consists of two
beamsplitters. Assume the absorbing medium is described by the
operator $S(ga + fb)$. We use Eq. (\ref{0004}), where the operator
of evolution has the form $T_{AB}=US(ga+fb)U^{\dagger}$. We take the
modes $a$ and $b$ in the coherent state $\ket{\alpha}$ and Fock
state $\ket{n}$ respectively. Let the modes $a$ and $b$ be mixed on
the first beamsplitter  and come to the absorber at whose output the
second beamsplitter is placed. According to Eq. (\ref{0004}), the
Fock state can be reproduced at one of the outputs of the scheme and
the coherent state, which is however weakened, can be found  at the
other output.
\\
To describe the light, we use the normally ordered characteristic
function $C_{N}$, which, in contrast to the Glauber-Sudarshan
quasiprobability $P$, is a nonsingular function for the Fock states.
The function $C_{N}$ is defined as the mean value of the
displacement operator being the Fourier transformation of the
Glauber-Sudarshan quasiprobability $P$:
\begin{eqnarray}
\nonumber
 C_{N}(\beta_{1},\beta_{2})= Sp\{\rho
D_{N}(\beta_{1})D_{N}(\beta_{2})\} &&\\
=\int d^{2}\alpha_{1} d^{2}\alpha_{2}P(\alpha_{1},\alpha_{2})e^
{\beta_{1}\alpha_{1}^{*}+\beta_{2}\alpha_{2}^{*}-
\beta_{1}^{*}\alpha_{1}-\beta_{2}^{*}\alpha_{2}},
\end{eqnarray}
where
$D_{N}(\beta_{k})=\exp(\beta_{k}c_{k}^{\dagger})\exp(-\beta_{k}^{*}c_{k}),
$ $k=1,2, c_{1}=a, c_{2}=b,$ and $\rho$ is the density matrix of the
modes $a$ and $b$. If initially the modes $a$ is in  coherent state
$\ket{\alpha}$ and $b$ is in the Fock state $\ket{n}$ then the
outgoing characteristic function of light has the form
\begin{eqnarray}
\label{0007}
 \nonumber
C_{N}(\beta_{1},\beta_{2})=
\nonumber
 \exp\{(c\beta_{1}-s\beta_{2})\alpha^{*}-
(c\beta_{1}-s\beta_{2})^{*}\alpha\}&&\\
\sum_{k=0}^{n}C^{n}_{k}\frac{(-1)^{k}}{k!}
|s\beta_{1}+c\beta_{2}|^{2k},
\end{eqnarray}
where $c,s$ are transmittance and reflectance of the beamsplitters.
\\
To describe propagating light trough the absorbing medium we will
use the master equation for the field. This equation can be achieved
with the help of adiabatic illumination procedure presented in
\cite{11} and the formalism of the quantum transfer theory
\cite{12}. For the simple case of the one dimensional problem in the
Fokker-Planck approximation, the master equation has the form
\begin{eqnarray}\label{00071}
\nonumber
 (\partial/\partial t+v\partial/\partial z)C(z,t)=&&\\
\nonumber R\Big[ ((c-sf/g)\partial/\partial h
+(s+cf/g)\partial/\partial
e)&&\\
(hc-hsf/g+se+ecf/g)+c.c.\Big]C(z,t),&&
\end{eqnarray}
where $R=g^{2}{N/\gamma}$ is an absorbtion coefficient, $N$ - is
occupation of the lover level of atom, $\gamma$ - is a transversal
decay rate and
\begin{eqnarray}
\label{000711} \nonumber
h=c\beta_{1}-s\beta_{2}&&\\
 e=s\beta_{1}+c\beta_{2}.
\end{eqnarray}
To solve Eq. (\ref{00071}) it needs boundary conditions, which can
be taken in the form (\ref{0007}), that corresponds to the light at
the output of the first beamsplitter.
\\
Let $sg=-cf$. Then all derivatives over $e$ vanish and the solution
has the form
\begin{eqnarray}
\label{00072} C_{N}(z) =
\exp[q(h\alpha^{*}-h^{*}\alpha)] 
\sum_{k=0}^{n}C^{n}_{k}\frac{(-1)^{k}}{k!} |e|^{2k},
\end{eqnarray}
where $q=\exp(-Mz)$, $M=Rc^{2}(1+(f/g)^{2})$.

 To achieve the state of light at the output of the absorber it needs
 changing the variables $\beta_{1}, \beta_{2}$ in (\ref{00072}) using
(\ref{000711}). The unitary transformation of light by the second
beamsplitter behind the absorber is described by replacing variables
given by (\ref{000711}). It has the form
$\beta_{1}'=c\beta_{1}-s\beta_{2},
\beta_{2}'=s\beta_{1}+c\beta_{2}$, where the beamsplitters are
assume to be identical. As a result the outgoing characteristic
function reads $C_{N}(\beta_{1}',\beta_{2}')=
\exp[q(\beta_{1}'\alpha^{*}-\beta_{1}'^{*}\alpha)]
\sum_{k=0}^{n}C^{n}_{k}(-1)^{k}|\beta_{2}'|^{2k}/k!$. It describes
two independent modes in the coherent state $\ket{q\alpha}$ and in
the Fock state $\ket{n}$ obtained by transformation
\begin{eqnarray}
 \ket{\alpha}\otimes\ket{n}\to \ket{\alpha  e^{-Mz}}\otimes\ket{n}.
\end{eqnarray}
It tells, that the initial Fock state is reproduced at the output of
the scheme and the amplitude of the coherent mode decreases.
\\
We are grateful to A.M. Basharov and S.P. Kulik for discussions.
This work is partially supported by the Delzell Foundation inc.
\section{appendix}
\appendix

We briefly discuss the properties of light obtained when the
coherent state, which is usually associated with the wave nature of
light, and the Fock state, which is associated with the corpuscular
nature of light, are mixed by the beamsplitter. The characteristic
function at the output of the beamsplitter, given by Eq.
(\ref{0007}), describes the pure state
\begin{eqnarray}
\label{p1}
A_{n\alpha}=(1/\sqrt{n!})(sa+cb)^{n}A_{0\alpha},&&\\
A_{0\alpha}=\ket{c\alpha}\otimes\ket{-s\alpha},&& \nonumber
\end{eqnarray}
where $\alpha$ and  $n$ are the complex amplitude of the coherent
mode and the number of photons of the Fock mode. Thanks to the Fock
state the nonclassical correlation of intensity of light aries and
the sub-Poissonian statistics of photons can be found. Next two
features are true: 1/ each of modes has sub-Poissonian statistics of
photons and the Mandel parameter of the mode $a$, for example, reads
$\xi_{a}=s^{2}n^{2}2c^{2}(|\alpha|^{2}-s^{2})/
(c^{2}|\alpha|^{2}+s^{2}n)\geq -1$;~~ 2/ the joint photon
coincidence count rate is lower than its value for a random flux
$\mean{a^{\dagger}ab^{\dagger}b}-\mean{a^{\dagger}a}\mean{b^{\dagger}b}
=-c^{2}s^{2}n (2|\alpha|^{2}+1)$;~~ 3/ the variance of the photon
number operator $u=a^{\dagger}a+\epsilon b^{\dagger}b$, where
$\epsilon=\pm 1$, can be lower than the shot level
$\mean{a^{\dagger}a+b^{\dagger}b}$ described the standart quantum
limit. It means the suppression of the shot noise when the sum or
difference of photocurrents of two detectors are measured.
\\
Another class of correlations that are responsible for squeezing and
entanglement is described by the quadrature operators of the modes,
given by the operators of the canonical position and momentum. These
correlations are measured in heterodyne schemes and they are phase
sensitive. Because of the Fock state, which has not phase in the
sense of the phase-space representation, the state $A_{n\alpha}$ is
not squeezed.
\\
Is the state $A_{n\alpha}$ entangled? It follows from (\ref{p1})
that the wave function is not factorized and in accordance with this
property the state $A_{n\alpha}$ is entangled. For $n=1$ the wave
function $A_{1\alpha}$ reads
$A_{1\alpha}=(1/\sqrt{2})(\ket{t_{0}t_{1}}-\ket{t_{1},t_{0}})$,
where $t_{m}=a^{\dagger m}\ket{\alpha/\sqrt{2}}$. It looks as EPR
pair  of discrete variables. In the same time because of the
coherent state we can try to analyze entaglement with respect to
continuous variables, using the cririon of non-separability
\cite{13}. According to this criterion for any separable state the
inequality
\begin{eqnarray}
  C=\mean{(\Delta Q)^{2}}+\mean{(\Delta P)^{2}}\geq 2
\end{eqnarray}
is valid, where the variances of the canonical operators of total
position $Q=x_{a}+x_{b}$ and relative momentum $P=p_{a}-p_{b}$ are
introduced and $a=x_{a}+ip_{a}, b=x_{b}+ip_{b}$. If this inequality
is invalid, the question of the separability of the state is open.
The reason is that this criterion is necessary and sufficient only
for Gaussian fields which the Wigner function has the Gaussian form.
In our case the Wigner function of $A_{n\alpha}$ is non-Gaussian,
but it is really doesn't matter, because the inequality is valid:
\begin{eqnarray}
C =2(1+n)\geq 2.
 \end{eqnarray}
Thus the state $A_{n\alpha}$ is separable or unentangled. We remain
open the  problem of the relations between the factorizability of
the wave function, entanglement and separability. However, the
presented example shows that the state $A_{n\alpha}$ has opposite
properties with respect to discrete and continuous variables.
Nevertheless, the state can be used, for example, as a quantum
channel for the standard protocol of teleportation of discrete
variables. To prove this statement, it is sufficient to note that
$A_{n\alpha}$ is unitary equivalent to a set of the Fock states
which is a complete basis represented a Bell-like state measurement.
Then new protocol is achieved by a simple local unitary
transformation and a set of the recovering operators can be found.
However, the separability of $A_{n\alpha}$ means that the state is
at least free of the property of an EPR pair of continuous variables
\cite{14} for which $C = 0$.

\end{document}